\documentstyle[prl,aps,epsf]{revtex}\tighten\twocolumn
\input epsf.sty
\begin{document}
\draft

\title{
Re-appearance of antiferromagnetic ordering with Zn and Ni
substitution in La$_{2-x}$Sr$_x$CuO$_4$
}

\author{
T. Machi,$^{1}$ 
I. Kato,$^{1,2}$ 
R. Hareyama,$^{1,2}$ 
N. Watanabe,$^{1,2}$ 
Y. Itoh,$^{1,3}$ 
N. Koshizuka,$^{1}$ 
S. Arai,$^{2}$ 
M. Murakami,$^{1}$ 
}

\address{
$^{1}$Superconductivity Research Laboratory,
 International Superconductivity Technology Center,
1-10-13 Shinonome, Koto-ku, Tokyo 135-0062, Japan,\\
$^{2}$Meiji University, 1-1-1 Hogashi-mita,
Tama-ku, Kawasaki-shi, Kanagawa 214-0033, Japan,\\
$^{3}$ Japan Society for the Promotion of Science, Tokyo, Japan
}
 
\date{\today}%
\maketitle %

\begin{abstract}
The effects of nonmagnetic Zn and magnetic Ni substitution for Cu site on
magnetism are studied by measurements of uniform magnetic susceptibility
for lightly doped La$_{2-x}$Sr$_x$Cu$_{1-z}$M$_z$O$_4$ (M=Zn or Ni)
polycrystalline samples. For
the parent $x$=0, Zn doping suppresses the N\'{e}el
temperature
$T\mathrm_N$ whereas Ni doping hardly changes $T\mathrm_N$ up to
$z$=0.3. For the lightly doped samples with $T\mathrm_N$$\sim$0, the Ni
doping  recovers $T\mathrm_N$. For the superconducting samples, the Ni
doping induces the superconductivity-to-antiferromagnetic transition (or
crossover). All the heavily Ni doped samples indicate a spin glass
behavior at
$\sim$15 K.  
\end{abstract}
\pacs{74.72.Dn, 75.40.Cx, 75.30.Hx}
%\end{minipage}} %

Although nonmagnetic impurity Zn substitution effect has been extensively
studied for high-$T\mathrm_c$ cuprate superconductors and the parent Mott
insulators, magnetic impurity Ni substitution effect has not
been extensively studied relatively.
Particularly, to our knowledge, there are a few studies for Ni doping
effect in semiconducting regime (~\cite{Fuji,Ohba,Ting,Sree}). 
In this
paper, we report a systematic study of Ni substitution effect on the
polycrystalline samples of La$_{2-x}$Sr$_x$Cu$_{1-z}$M$_z$O$_4$ in the
parent antiferromagnet, the lightly doped insulators without long range
order, and the relatively low-$T\mathrm_c$ superconductors, through
measurement of uniform magnetic susceptibility $\chi$. 
The polycrystalline samples
were synthesized by a solid state reaction method. 
For comparison, we
synthesized also La$_{2-x}$Sr$_x$Cu$_{1-z}$Zn$_z$O$_4$
\cite{Hucker}. Here, we emphasize an importance of careful annealing
process at 650 $\mathrm^o$C for 48 hours under Ar gas atmosphere. The
uniform magnetic susceptibility was measured by a SQUID magnetometer. 
The N\'{e}el
temperature $T\mathrm_N$ is determined by the maximum behavior, or the
onset temperature of hysteresis of the magnetic susceptibility between zero
field cooling (ZFC) and field cooling (FC). The spin glass temperature
$T\mathrm_{SG}$ is defined by the low temperature sharp peak in the further
hysteresis \cite{Chou}. For non-superconducting samples, a magnetic field
of 100$\sim$1.0$\times$10$^4$ Oe was applied, whereas for superconducting
samples, a field of
$\sim$ 100 Oe was applied.  

Figure 1 shows Ni doping effect on the $T$ dependence
of magnetic susceptibility. We found the followings: 

1) Up to $z$=0.3 for pure La$_2$CuO$_4$, Ni doping does not destroy
the N\'{e}el ordering. 
Such a robust $T\mathrm_N$ to Ni doping is in contrast to a fragile
$T\mathrm_N$ to Zn doping \cite{Uchinokura,Ting}. In Fig, 2, for
comparison, $T\mathrm_N$ versus Ni or Zn content $z$ is shown. 

2) With further Ni doping $z>$0.3, the spin glass ordering appears at
$T\mathrm_{SG}\sim$15 K, probably due to Ni spin freezing. Hereafter, we
call this Ni freezing temperature.  

3) The N\'{e}el ordering, which is suppressed down to $T\mathrm_N<$4.2 K
by Sr doping $x$=0.02, recovers more rapidly and largely with Ni doping,
than with Zn doping ~\cite{Hucker}.  

4) The superconductivity for Sr $x$=0.06 or 0.08 is easily suppressed
by Ni doping. The Ni doping induces the
superconductor-to-antiferromagnet transition (crossover) at
$z$=0.02$\sim$0.04. Further Ni doping for $z>$0.3 or $>$0.2 induces the
spin glass state with $T\mathrm_{SG}\sim$15 K. 

5) The Ni freezing temperature with heavily Ni doping does
not seem to depend on Sr doping level.  

6) The Ni freezing temperature $T\mathrm_{SG}\sim$15 K is about two times
larger than the Ni-free, spin glass temperature $T\mathrm_g\sim$7 K at
Sr
$x$=0.04 ~\cite{Chou}.

In Fig. 3, we summarize the magnetic phase diagram versus Ni content
$z$ at various Sr doping, which can be drawn from the preset study in
Fig. 1. The magnetic impurity Ni doping yields rich phases through the
order-to-disorder transition (or crossover) in antiferromagnetic
correlation, the spin glass transition on Ni spin freezing, and the
superconductor-to-insulator transition (or crossover).  In conclusion, we
demonstrate that Ni doping causes the above novel effects on the strongly
correlated electron system La$_{2-x}$Sr$_x$CuO$_4$.  

This work was supported by NEDO.
 
\begin{figure}
\epsfxsize=3.7in
\epsfbox{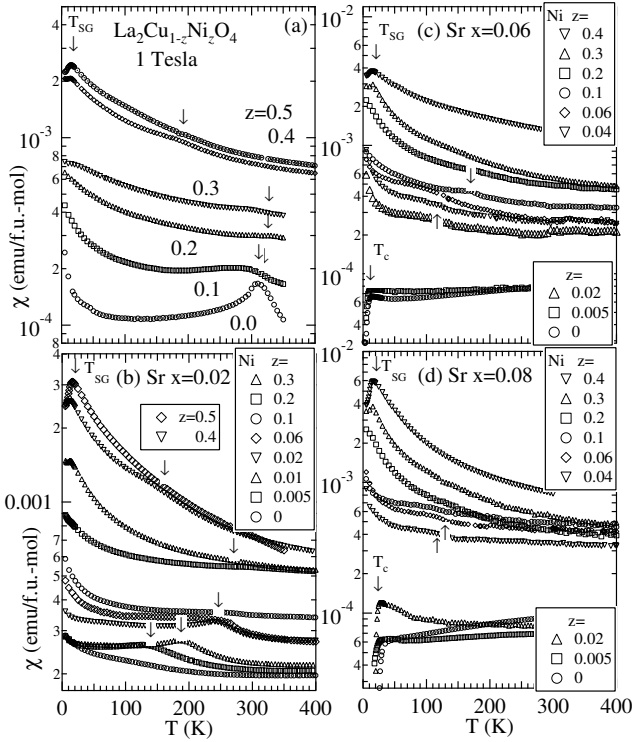}
\vspace{0.2cm}
\caption{
Ni-doping effect on the ZFC $dc$ magnetic susceptibility of
La$_{2-x}$Sr$_x$Cu$_{1-z}$M$_z$O$_4$; parent insulator $x$=0 (a),
lightly doped non-superconducting $x$=0.02 without long range order at
$z$=0 (b), lightly doped $x$=0.06 with relatively
low $T\mathrm_c$ at $z$=0 (c), and $x$=0.08 at $z$=0 (d).
The arrows without character indicate $T\mathrm_N$'s, the other arrows
with character are $T\mathrm_c$'s, or $T\mathrm_{SG}$'s. For simplicity,
we do not attach all the arrows.
}
\label{Suscep}
\end{figure}

\begin{figure}
\epsfxsize=3.8in
\epsfbox{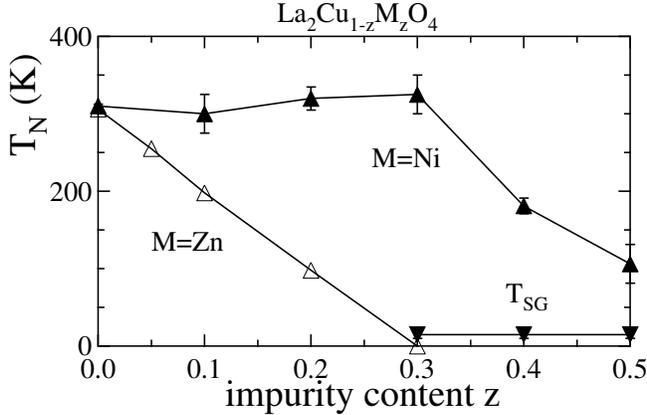}
\vspace{0.2cm}
\caption{
La$_2$Cu$_{1-z}$M$_z$O$_4$: $T\mathrm_N$ versus M=Ni or Zn content $z$.
The result is qualitatively consistent with Refs. [7, 3].  
}
\label{ZnvsNi}
\end{figure}

\begin{figure}
\epsfxsize=3.8in
\epsfbox{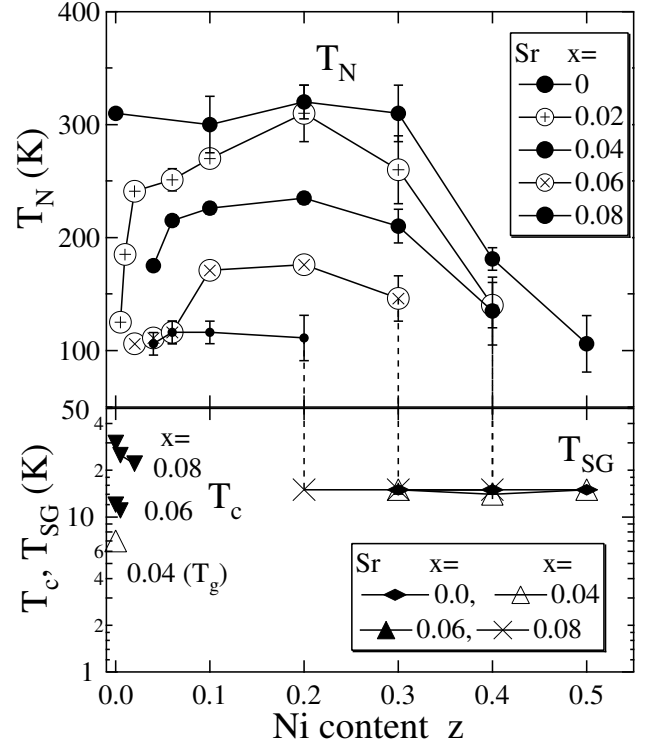}
\vspace{0.2cm}
\caption{
Magnetic phase diagram of La$_{2-x}$Sr$_x$Cu$_{1-z}$M$_z$O$_4$;
$T\mathrm_N$, $T\mathrm_{SG}$, $T\mathrm_g$, and $T\mathrm_c$ versus
Ni content $z$ at Sr doping. The solid and the dashed lines are guide for
the eye.   
}
\label{Diagram}
\end{figure}

% references


\begin{references} 

\bibitem{Fuji}H. Fujishita, M. Sato, Solid State Commun. {\bf 72} (1989) 529.

\bibitem{Ohba}K. Ohbayashi, H. Tsukamoto, H. Yamashita, A. Fukumoto, Y. Utsunomiya, N. Ogita, M. Udagawa, S. Funahashi, J. Phys. Soc. Jpn. {\bf 59} (1990) 1372.

\bibitem{Ting}S. T. Ting, P. Pernambuco-Wise, J. E. Crow, E. Manousakis, J. Weaver, Phys. Rev. B 46 (1992) 11772.

\bibitem{Sree}K. Sreedhar, P. A. Metcalf, J. M. Honig, Physica C {\bf 227} (1994) 160. 

\bibitem{Hucker}M. H\"{u}cker, V. Kataev, J. Pommer, J. Harra$\beta$, A. Hosni, C. Pflitsch, R. Gross, B. B\"{u}chner, Phys. Rev. B {\bf 59} (1999) R725.

\bibitem{Chou}F. C. Chou, N. R. Belk, M. A. Kastner, R. J. Birgeneau, A. Aharony, Phys. Rev. Lett. {\bf 75} (1995) 2204.

\bibitem{Uchinokura}K. Uchinokura, T. Ino, I. Terasaki, I. Tsukada, Physica B {\bf 205} (1995) 234.
  
\end{references}
\end{document}